\documentclass[longauth,traditabstract]{aa}

\newcommand{\mstar}{M$_\odot$}
\newcommand{\IRAS}{{\it IRAS}}

\newcommand{\Planck}{{\it Planck}}
 
\newcommand{\microm}{$\mu$m}

\newcommand{\disperse}{DisPerSE}

\newcommand{\hii}{{H{\scriptsize II}}}
\newcommand{\herschel}{{\it Herschel}}
\newcommand{\Av}{{$A_{\mathrm{V}}$}}

\newcommand{\ch}[1]{{#1}}
\newcommand{\cha}[1]{{#1}}

\usepackage{natbib}
\bibpunct{(}{)}{;}{a}{}{,} \usepackage{graphicx}
\usepackage{wasysym}
\usepackage{rotating}
\usepackage{pstricks}
\usepackage{txfonts}

\begin{document}
   \title{Filaments and ridges in Vela~C revealed by \herschel\thanks{\herschel\ is an ESA space observatory with science instruments provided by European-led Principal Investigator consortia and with important participation from NASA.}: from low-mass to high-mass star-forming sites\thanks{Fig. \ref{fig:hipe} is only available in electronic form at http://www.aanda.org.}}

   \author{T. Hill
          \inst{1}
          \and
          F. Motte\inst{1}
          \and
          P. Didelon\inst{1}
          \and
          S. Bontemps\inst{2}
          \and
          V. Minier\inst{1}
          \and
          M. Hennemann\inst{1}
          \and
          N. Schneider\inst{1}
          \and
          Ph. Andr\'e\inst{1}
          \and
          A.~Men`shchikov\inst{1} 
          \and
           L.~D.~Anderson\inst{3}
          \and
          D. Arzoumanian\inst{1}
          \and
          J.-P. Bernard\inst{4}
          \and
          J. di Francesco\inst{5}
          \and
          D. Elia\inst{6}
          \and
          T. Giannini\inst{7}
          \and
          M.~J.~Griffin\inst{8}
                              \and
          V. K\"onyves\inst{1}
          \and
          J. Kirk\inst{8}
          \and
          A. P. Marston\inst{9}
          \and       
          P. G. Martin\inst{10}  
          \and
          S. Molinari\inst{6}
          \and
          Q. Nguy$\tilde{\hat{\rm e}}$n Lu{\hskip-0.65mm\small'{}\hskip-0.5mm}o{\hskip-0.65mm\small'{}\hskip-0.5mm}ng\inst{1}
          \and
          N. Peretto\inst{1}
          \and
          S.~Pezzuto\inst{6}
          \and
          H. Roussel\inst{11}
          \and
          M.~Sauvage\inst{1}
          \and
          T.~Sousbie\inst{11}
          \and
          L. Testi\inst{12}
          \and
          D. Ward-Thompson\inst{8}
          \and
          G. J. White\inst{13,14}
          \and
          C.~D.~Wilson\inst{15}
          \and
          A.~Zavagno\inst{3}
                }
   \institute{Laboratoire AIM, CEA/IRFU CNRS/INSU Universit\'e Paris Diderot, CEA-Saclay, 91191 Gif-sur-Yvette Cedex, France\\
              \email{tracey.hill@cea.fr}
         \and 
         Universit\'e de Bordeaux, OASU, Bordeaux, France
         \and
         Laboratoire d’Astrophysique de Marseille UMR6110, CNRS, Universit\'e de Provence, 38 rue F. Joliot-Curie, 13388 Marseille, France
         \and
         Universit\'e de Toulouse, UPS, CESR, 9 avenue du colonel Roche, 31028 Toulouse Cedex 4, France ; CNRS, UMR5187, 31028 Toulouse
         \and
         National Research Council of Canada, Herzberg Institute of Astrophysics, University of Victoria, Department of Physics and Astronomy, Victoria, Canada
         \and
         INAF-Istituto Fisica Spazio Interplanetario, via Fosso del Cavaliere 100, 00133 Roma, Italy 
         \and
         INAF- Osservatorio Astronomico di Roma, via di Frascati 33, 00040
         Monte Porzio Catone, Italy.
         \and
         School of Physics and Astronomy, Cardiff University, Queens Buildings, The Parade, Cardiff, CF243AA, UK       
         \and
         Herschel Science Centre, ESAC, Spain
         \and
         Canadian Institute for Theoretical Astrophysics, University of Toronto, Toronto, ON, M5S 3H8, Canada
         \and
         Institut d’Astrophysique de Paris, and Universit\'e Pierre et Marie Curie (UPMC), UMR 7095 CNRS, 98 bis boulevard Arago, 75014
         \and
         ESO, Karl Schwarzschild str. 2, 85748 Garching bei Munchen, Germany
         \and
         The Rutherford Appleton Laboratory, Chilton, Didcot, OX11 0NL, UK
         \and
         Department of Physics and Astronomy, The Open University, Milton Keynes, UK
         \and
         Department of Physics and Astronomy, McMaster University, Hamilton, ON, L8S 4M1, Canada
         \email{}
             \thanks{}
             }

   \date{29 July 2011}

 \abstract{We present the first \herschel\ PACS and SPIRE results of the Vela~C molecular complex in the far-infrared and submillimetre regimes at 70, 160, 250, 350, and 500\,\microm, spanning the peak of emission of cold prestellar or protostellar cores. Column density and multi-resolution analysis (MRA) differentiates the Vela~C complex into five distinct sub-regions.
Each sub-region displays differences in their column density and temperature probability distribution functions (PDFs), in particular, the PDFs of the `Centre-Ridge' and `South-Nest' sub-regions appear in stark contrast to each other. The Centre-Ridge displays a  bimodal temperature PDF representative of hot gas surrounding the \hii\, region RCW\,36 and the cold neighbouring filaments, whilst the South-Nest is dominated by cold filamentary structure. The column density PDF of the Centre-Ridge is flatter than the South-Nest, with a high column density tail, consistent with formation through large-scale flows, and regulation by self-gravity. At small to intermediate scales MRA indicates the Centre-Ridge to be twice as concentrated as the South-Nest, whilst on larger scales, a greater portion of the gas in the South-Nest is dominated by turbulence than in the Centre-Ridge. 
In Vela C, high-mass stars appear to be \ch{preferentially} forming in ridges, i.e.,  dominant high column density filaments.
}


   \maketitle

\section{Introduction}

Though relatively rare, high-mass ($>$\,8\mstar) stars play a major role in the energy budget of the interstellar medium where they contribute to shape the composition and evolution of their natal neighbourhood, as well as their host galaxy. The earliest stages of high-mass star formation remain a process that is not well understood, due to a lack of suitable high-mass star progenitors. \ch{Recent work has focused on finding candidate early stage protostars indicative of the earliest stages of high-mass star formation \citep[e.g.][]{beuther02, hill05, motte07, rathborne07}.}

The \herschel\ imaging survey of OB young stellar objects (HOBYS) 
capitalises on the unprecedented mapping capabilities of the \herschel\ space observatory \citep{griffin10} to compile the first complete, systematic and unbiased sample of nearby 
($<$\,3\,kpc) high-mass star progenitors \citep{motte10, schneider10, hennemann10, difrancesco10}. The HOBYS programme targets ten molecular complexes forming OB-type stars, one of which is Vela~C,  at five wavebands from 70 
to 500\,\microm. 
\herschel\ has already proven 
useful in tracing both high- and low-mass star formation at all evolutionary stages including  prestellar cores, protostars, and filaments, as well as more evolved objects like \hii\ regions \citep[e.g.][]{motte10, andre10, molinari10, zavagno10}.

\begin{figure*}[!t]
\vspace{-2cm}
\begin{center}
\includegraphics[height=\hsize, angle=270]{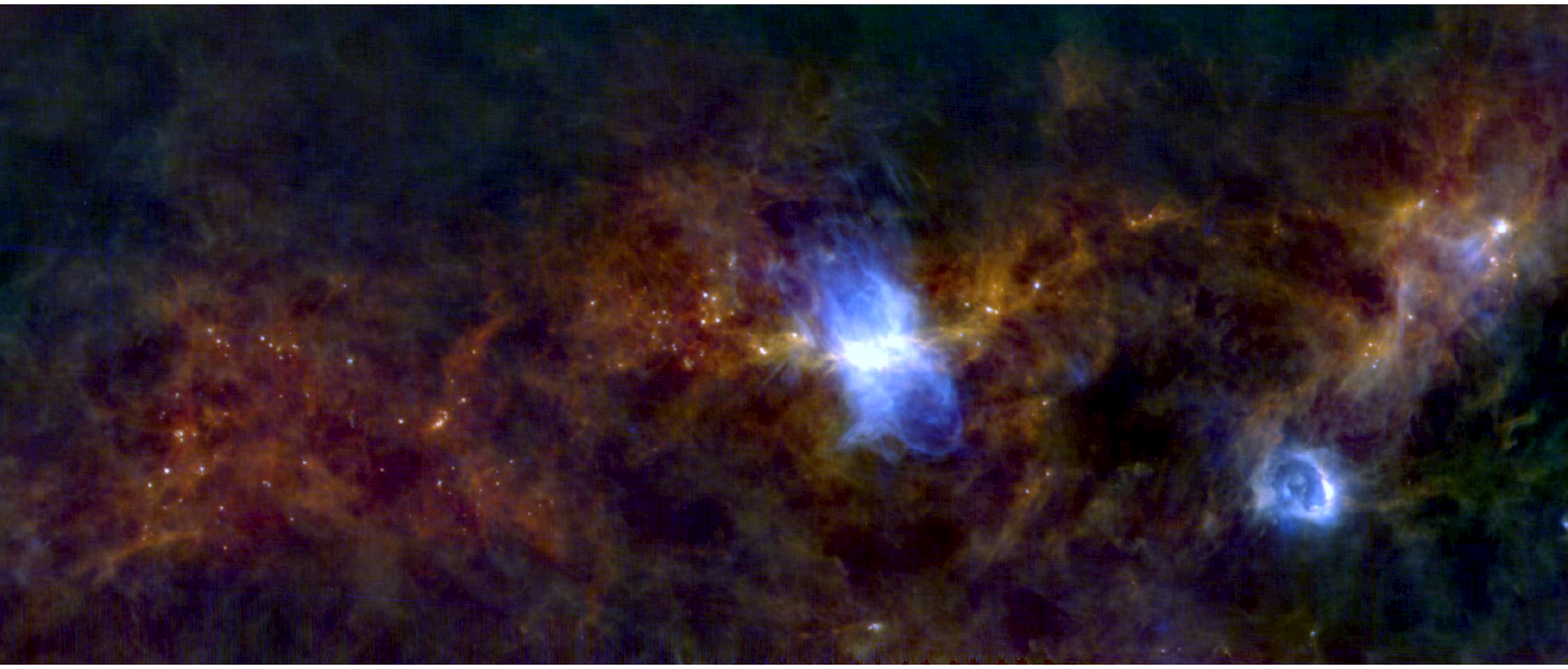} \vspace{-2.5cm}
\caption{3-colour image of Vela~C, red = 250\,\microm, green = 160\,\microm\ and blue = 70\,\microm. The shorter wavelengths reveal the hot dust, such as the  \hii\ regions RCW\,36 (centre) and RCW\,34 (right) which are clearly shown here in blue. The longer wavelengths show the cold, dense objects, such as the cold dense network of filaments  shown in red. There are many embedded sources within the filaments in Vela~C. A scale bar is given in Fig. \ref{fig:dusttemp} and the coordinates in Fig. \ref{fig:hipe}.
 \label{fig:3col} }
\end{center}
\end{figure*}

The Vela molecular region, often called the `Vela Molecular Ridge' in the literature 
\citep[][]{murphy91}, is a nearby giant molecular cloud complex within the Galactic plane that is comprised of four components labelled A through D.
The entire complex was observed in \ch{low resolution (0.5$^{\circ}$)} {\it J = 1-0} CO   by \citet{may88}, \ch{who found intense emission at relatively low radial velocities}. They 
estimated a total molecular
mass in the Vela region exceeding 5~$\times$~10$^5$\,\mstar, which was later confirmed by \citet{yamag99} with \ch{low resolution (8\,arcmin)} {\it J = 1-0} $^{12}$CO, $^{13}$CO, and C$^{18}$O. From \IRAS\ point sources and protostellar energy distributions, \citet{liseau92} suggested that tens of young massive stars exist toward the Vela region. The Balloon-borne Large Aperture Submillimetre Telescope \citep[BLAST;][]{netterfield09}, \ch{observing at three submillimetre wavelengths (250, 350, 500\,\microm) with resolutions between 36\arcsec\ and 60\arcsec,} confirmed that there are a large number of high-mass star formation sites in the Vela  molecular complex, including Vela C.

The Vela~C giant molecular cloud complex, the most massive component of the Vela region, is an ideal laboratory in which to study star formation. At a distance of 700 pc, Vela~C is a nearby star-forming complex at an early stage of star formation ($<$~10$^6$ years), containing molecular outflows and many star-forming regions detected in the far-infrared \citep[\IRAS;][]{wouterloot89, yamag99}. Vela~C is particularly unusual owing to its proximity {\it and} the fact that it houses  high, intermediate and low mass star formation \citep[e.g.][]{massi03} and so may provide clues as to what causes the different modes of star formation.

\section{Observations, data reduction, source extraction}

Vela~C was observed  on 2010, May 18, as part of the HOBYS guaranteed time key programme. The parallel-scan mode of \herschel\ was used, allowing simultaneous observations with the PACS \citep[70 and 160\,\microm;][]{pog10}  and SPIRE \citep[250, 350, 500\,\microm;][]{griffin10} instruments at five bands, using the slow scan-speed (20\arcsec/s). A total area of $\sim$~3~deg$^2$ was mapped \ch{using two orthogonal scan directions}. The PACS and SPIRE data were reduced with version 5.0.1975 of \cha{the Herschel Interactive Processing Environment \citep[HIPE\footnote{http://herschel.esac.esa.int/HIPE\_download.shtml};][]{ott10}}
 adopting standard steps of the default pipeline to level-1 including calibration and deglitching. \ch{The pipeline was modified, for both PACS and SPIRE, to include data taken during the turn-around of the telescope, allowing better baseline subtraction.}
\ch{Calibration of the PACS data has been found to be within 10\% and 20\% at 70 and 160\,\microm, respectively (see the PACS observers' manual), whilst SPIRE calibration is within 10\% for all bands (see the SPIRE observers' manual).}
Maps were produced using the HIPE level-1 data and v7 of the {\it Scanamorphos} software package\footnote{http://www2.iap.fr/users/roussel/herschel/index.html} which performs baseline and drift removal before regriding (Roussel et al., 2011, submitted). Individual images at each waveband are shown in Fig. \ref{fig:hipe}.
Sources were identified using the multi-wavelength multi-resolution {\it getsources} extraction routine \citep{mensh10} which is consistent with the 
MRE-GCL algorithm \citep{motte03} used for the HOBYS programme during the \herschel\ science demonstration phase \citep[e.g.][]{motte10}.

\section{Structure analysis}

The \herschel\ maps of Vela~C (Figs \ref{fig:3col}, \ref{fig:hipe}), show cold filamentary structures throughout the entire complex, as well as bright emission at the shorter wavelengths, associated with the RCW\,36 cluster at the centre and RCW\,34 in the north. RCW\,36 is powered by at least one O8 or two O9 stars \citep{baba04}.

\vspace{-0.2cm}
\subsection{Column density \& temperature maps}\label{sec:colden}

\begin{figure*}[]
\includegraphics[height=0.275\hsize]{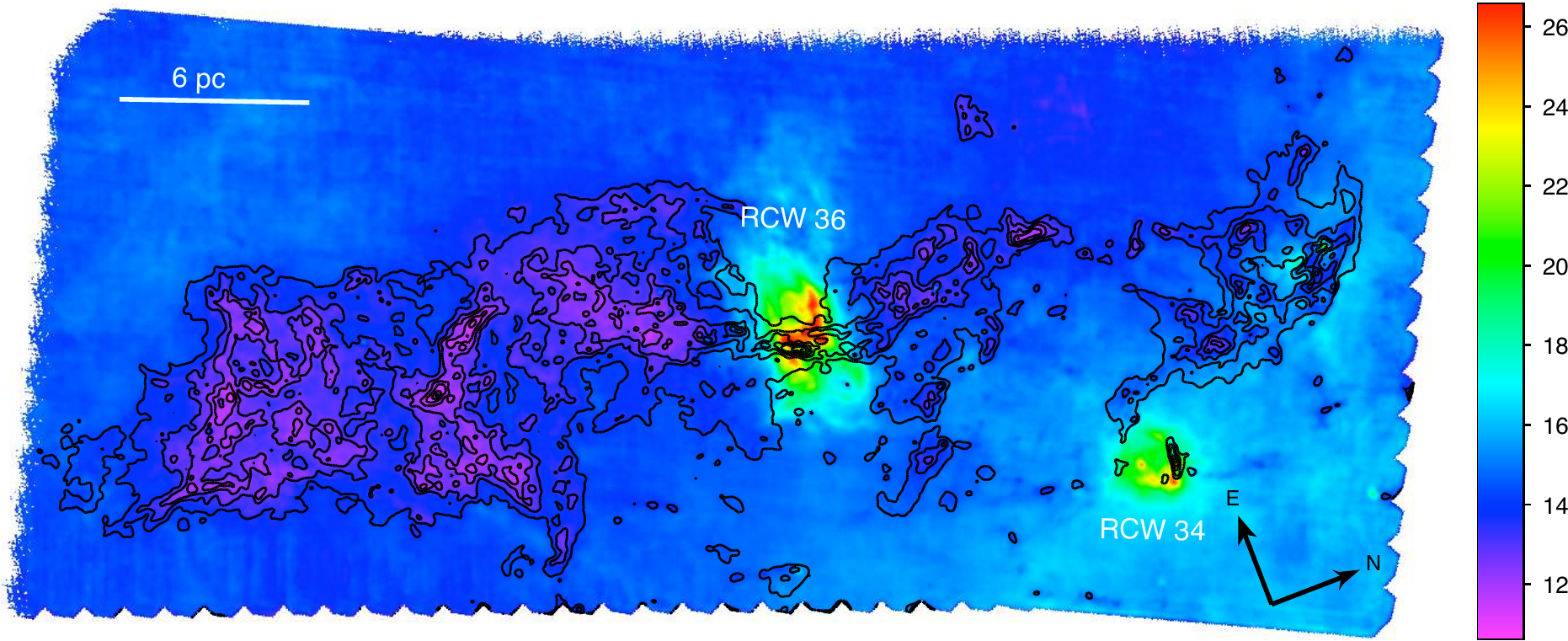}\hfill 
\includegraphics[height=0.264\hsize]{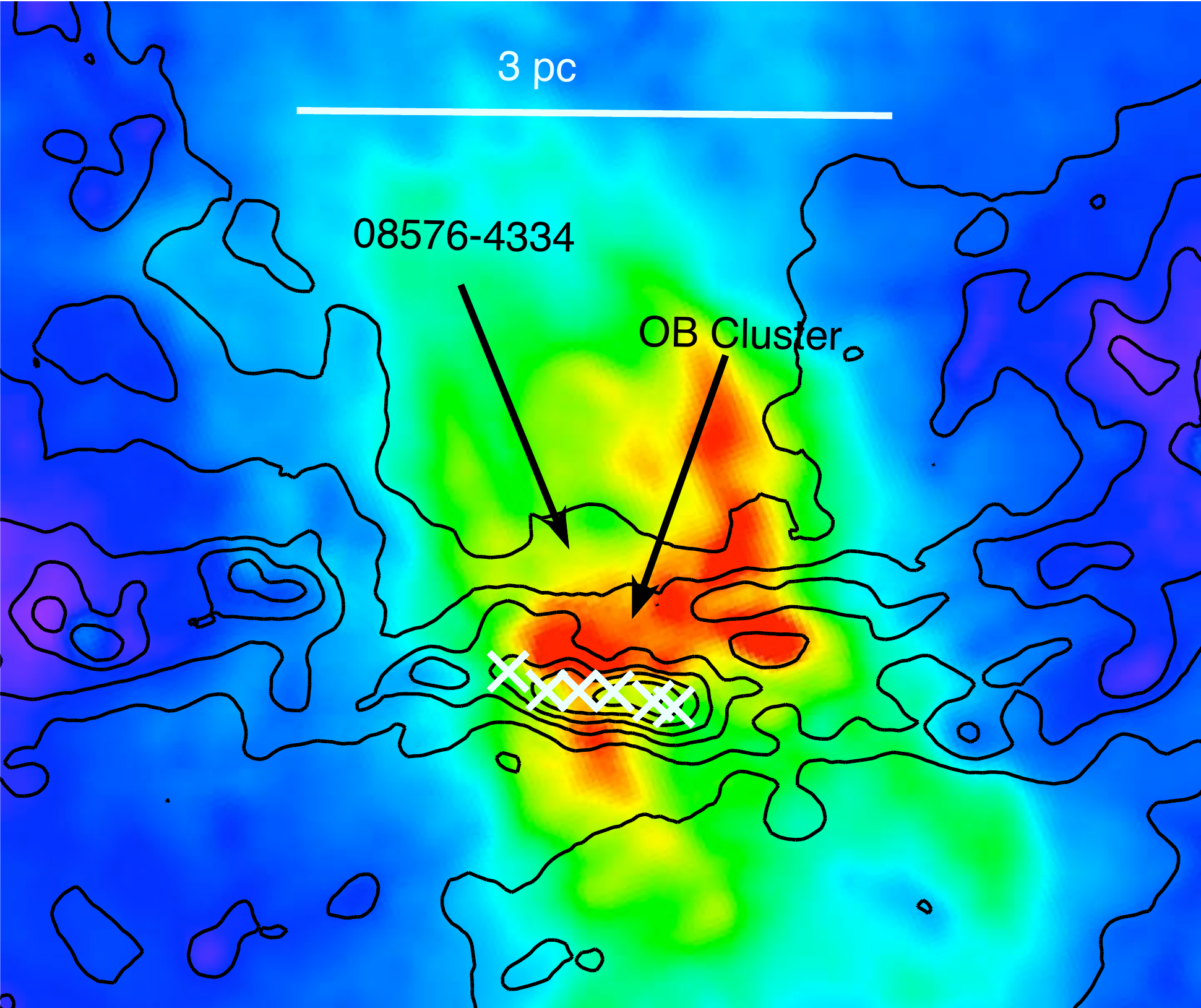}
\caption{Left: Dust Temperature map (37\arcsec, 0.12\,pc) of Vela~C, with the column density contours (Fig. \ref{fig:colden}) overlaid. The scale bar is 0.49 degrees.  
Contours are 9.0$\times$10$^{21}$, 1.5, 2.4, 3.9, 6.4$\times$10$^{22}$, 1.0$\times$10$^{23}$ cm$^{-2}$.
 Right: zoom to centre of map where a cold ridge is overlaid on the RCW\,36 region. White cross-marks correspond to those in black in Fig. \ref{fig:colden}. \label{fig:dusttemp}}
\end{figure*}  

\begin{figure*}
\includegraphics[height=0.275\hsize]{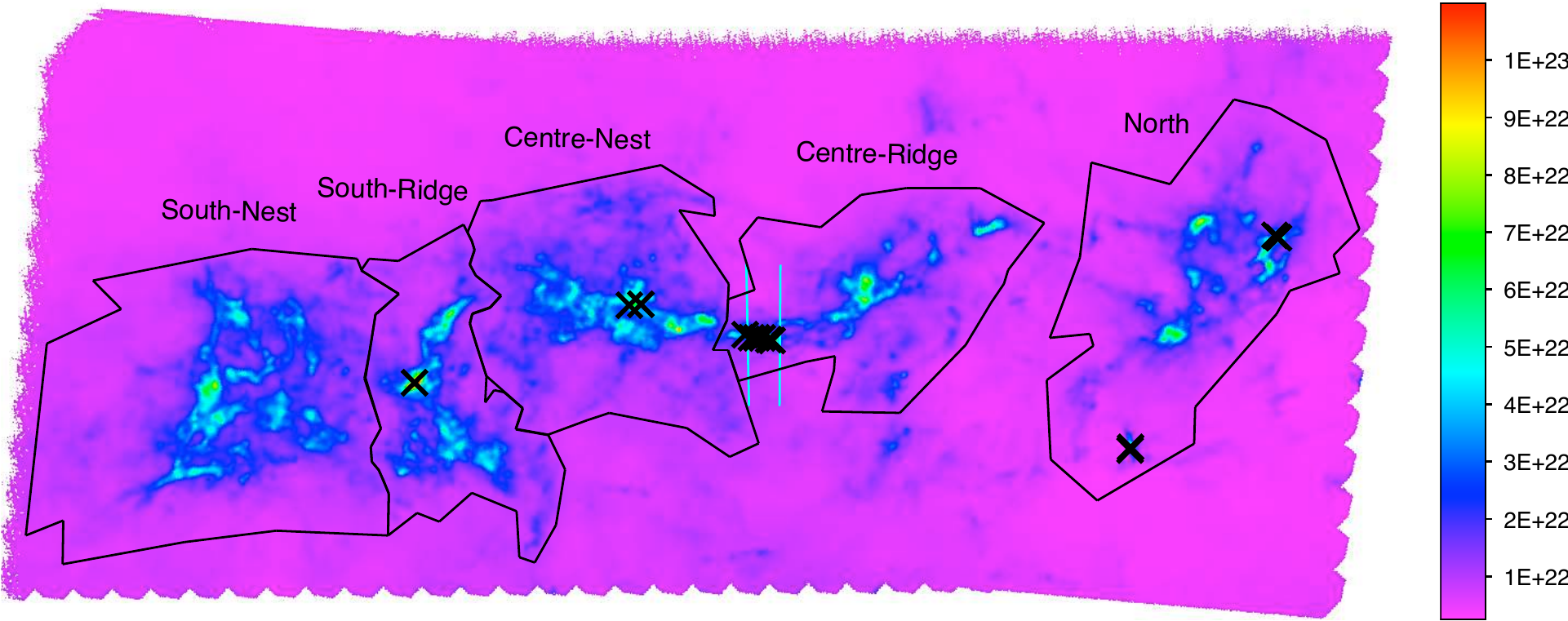}\hfill
\includegraphics[height=0.264\hsize]{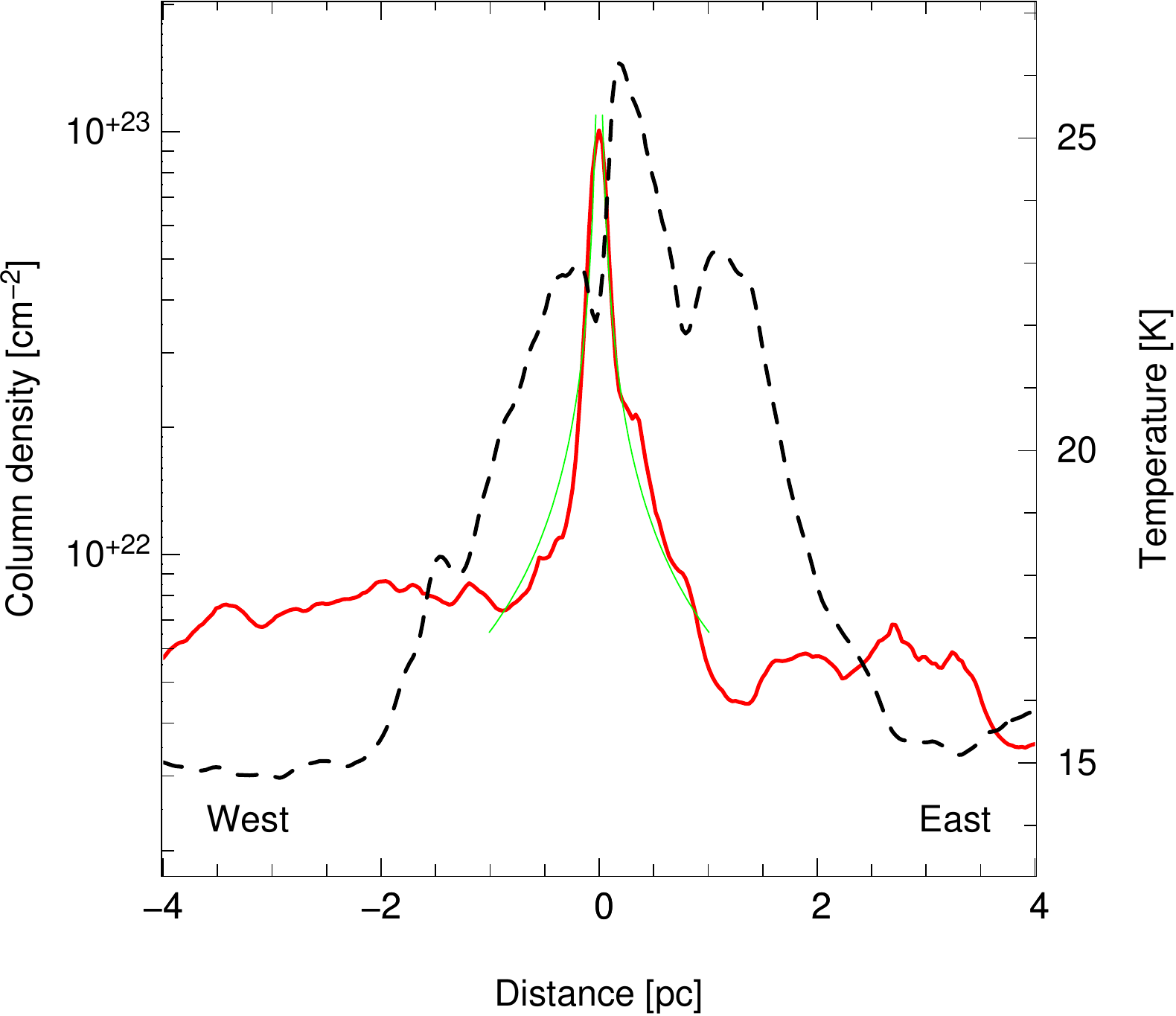}
\caption{Left: Column Density map with sub-regions, as defined at an \Av\ $>$ 7 mag, overlaid. The 13 most massive sources with (S:N\,$>$\,50) are denoted by a black cross. \cha{These sources have masses ranging from 20 -- 60\,\mstar.}
Right: Column density (red) and temperature (dashed black)  cut perpendicular to the densest part of the main filament (cyan lines on left image). The green curve is a power law fit N$_{H_2} \propto$~ r$^{-0.8}$  which is consistent with a $\rho$(r) $\propto$ r$^{-2}$ perpendicular to the filament. \label{fig:colden}}
\end{figure*}

The dust temperature and column density maps (Figs \ref{fig:dusttemp} and \ref{fig:colden}) of Vela C were drawn by fitting pixel-by-pixel spectral energy distributions \citep[SEDs; as described by][]{hill09, hill10} to the four longer \herschel\ wavebands (at  the same 37\arcsec\ resolution), assuming the dust opacity law  of \citet{hildebrand83}\ \ch{and a spectral index of 2. Only the four longer wavelength \herschel\ bands were used for SED fitting, as the 70\,\microm\ data may not be entirely tracing the cold dust which we are interested in.}

The zero offsets were determined \ch{following the procedure described in detail by \citet{bernard10}, and applied to the \herschel\ maps prior to fitting. This method uses the IRIS  (improved reprocessing of the \IRAS\ survey) data \citep{md05}, which is at comparable resolution to that of \Planck, as well as the \Planck\ HFI DR2 data \citep{plancka11} to predict the expected brightness in the \herschel\ PACS and SPIRE bands. The derived offsets are obtained from the correlation between the \Planck/\IRAS\ expectation and the actual \herschel\ data smoothed to the common resolution of the \IRAS\ and \Planck\ high frequency data.  This application to \herschel\ data has been used by \citet{bernard10,juvela11}.
An alternate method was used by \citet{konyves10} and \citet{schneider10} who used extinction maps to determine absolute calibration of their \herschel\ Aquila and Rosette column density maps. Without the proper absolute calibration, the dust temperature map can be overestimated by $\sim$10 to 15\%. The observed rms of cirrus noise determined directly from the column density map is $\sim$ 3.5\,$\times$\,10$^{21}$\,cm$^{-2}$, consistent with that found for Aquila \citep{andre10}.}
A lack of emission at \IRAS\ 100\,\microm, meant that the zero offsets could not be determined for the 70\,\microm\ band, confirming our earlier decision to exclude it from the dust temperature and column density maps.

These column density and temperature maps indicate that the southern part of Vela~C is dominated by cold, dense material, in contrast to the centre of the map  where a cold dense filament ($\sim$15--18\,K), as well as warmer and less dense material coexist.

Recent \herschel\ results of the Gould Belt survey key programme
have revealed the presence of star-forming filamentary structures
above an \Av\ $>$\,7 mag \citep{andre11,andre10}, which these authors
interpret as being `supercritical' (i.e., gravitationally unstable)
filaments forming prestellar cores in their interiors by gravitational
instability.
 At an \Av\ magnitude\footnote{Here we define \cha{N$_{H_2}$ =  1\,\Av\ $\times$ 10$^{21}$ cm$^{-2}$ mag$^{-1}$} \citep{bohlin78}.} of 7, the Vela~C complex segregates into five distinct sub-regions: North, Centre-Ridge, Centre-Nest,  South-Ridge, South-Nest.  (see Fig. \ref{fig:colden}). The cloud structures which have 2\,pc characteristic size-scales
 supports this segregation of Vela~C into five distinct sub-regions as shown by a multi-resolution analysis of the column density map (see Fig. \ref{fig:mrfilter} and section \ref{sec:disperse}). \ch{These sub-regions contain 1.9 -- 4.2\,$\times$\,10$^6$\,\mstar\ (see Table \ref{tab:crest}).}

\ch{The segregation of Vela C into five sub-regions is also seen in molecular line data, such as the CO and C$^{18}$O observations of \citet{yamag99}. In their Figure 4a, Vela~C clearly segregates into the North, South-Nest and South-Ridge sub-regions. The distinction between Centre-Ridge and Centre-Nest is not as clear but exists, even at the 8\arcmin\ resolution of these observations. There is a small velocity gradient throughout the Vela C molecular cloud \citep[see,][for more detail]{yamag99}.}

\begin{figure*}
\begin{center}
\includegraphics[width=15cm]{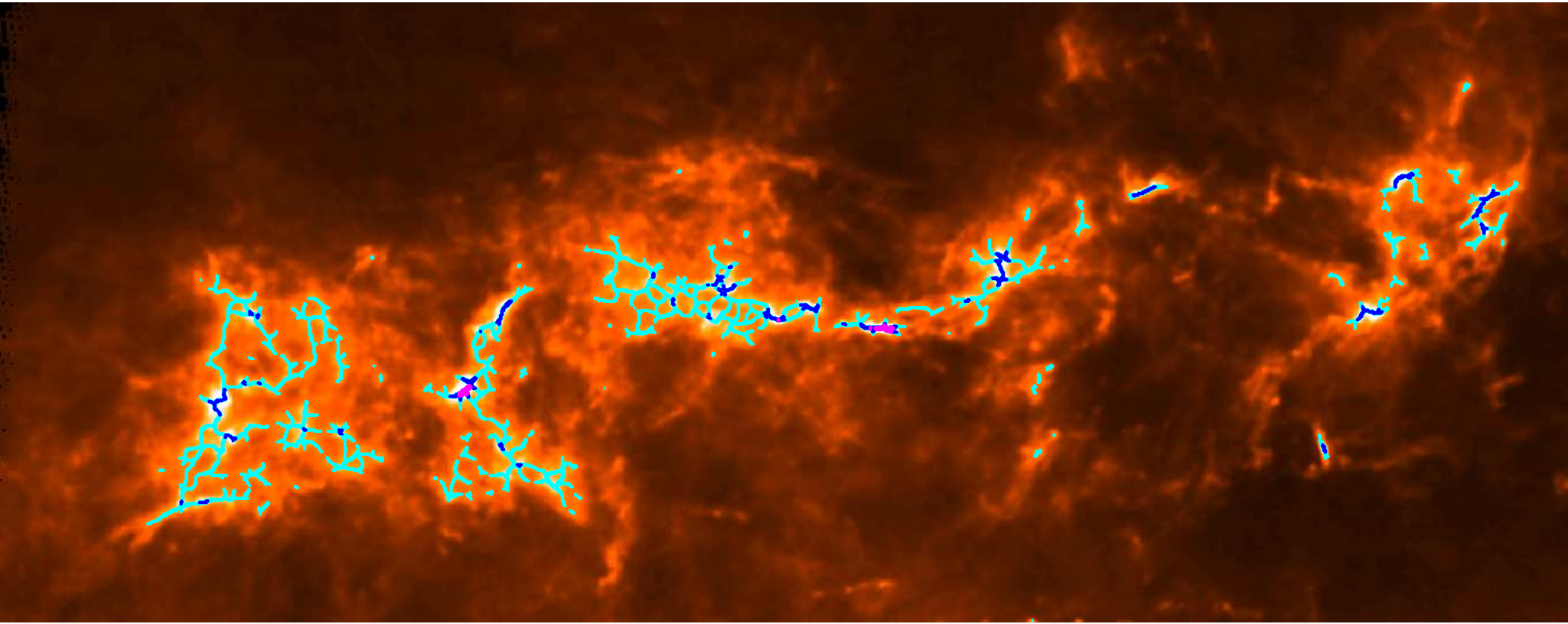}\\


\includegraphics[width=5.5cm]{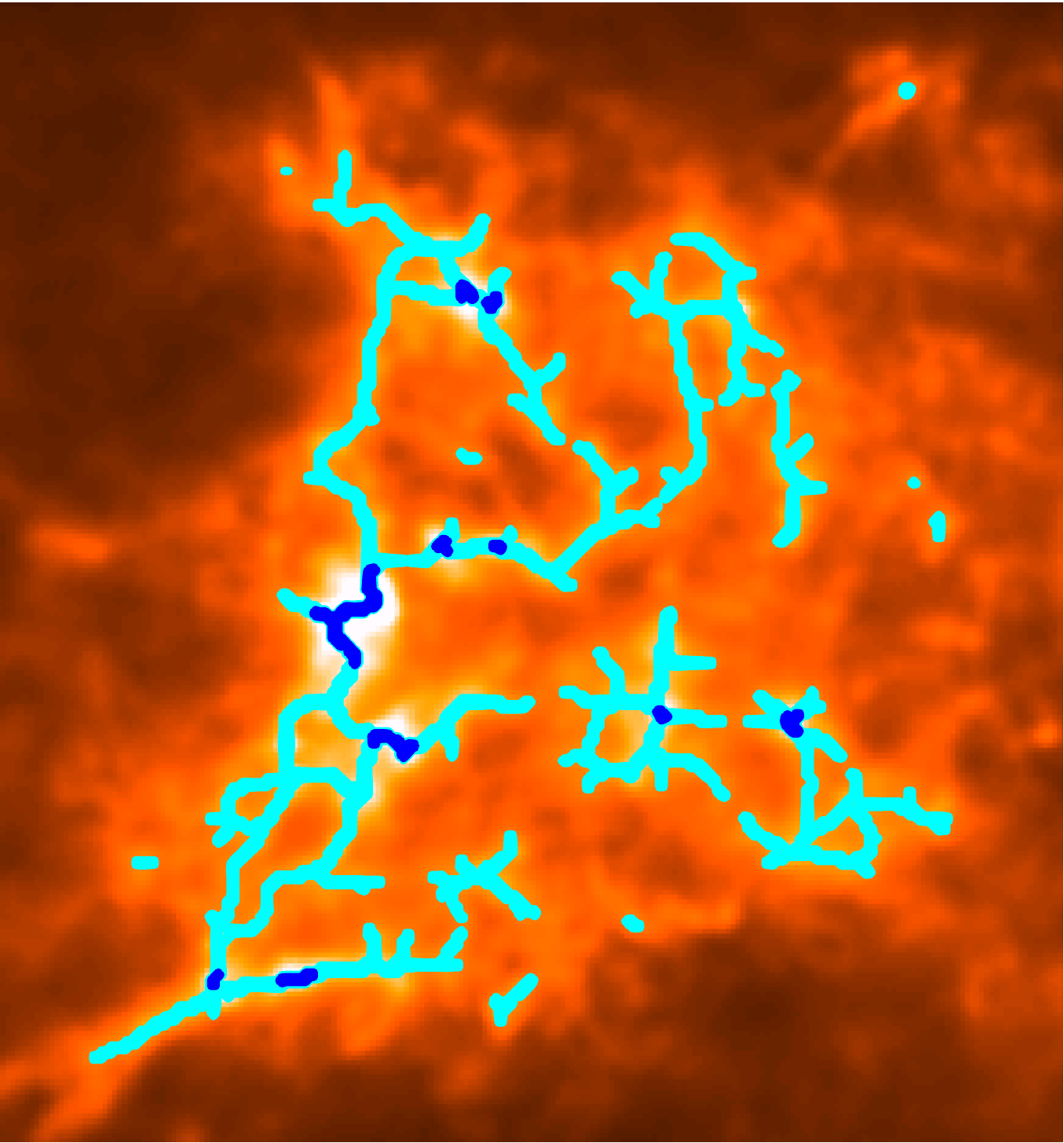}
\hspace{1cm}
\includegraphics[width=7.5cm]{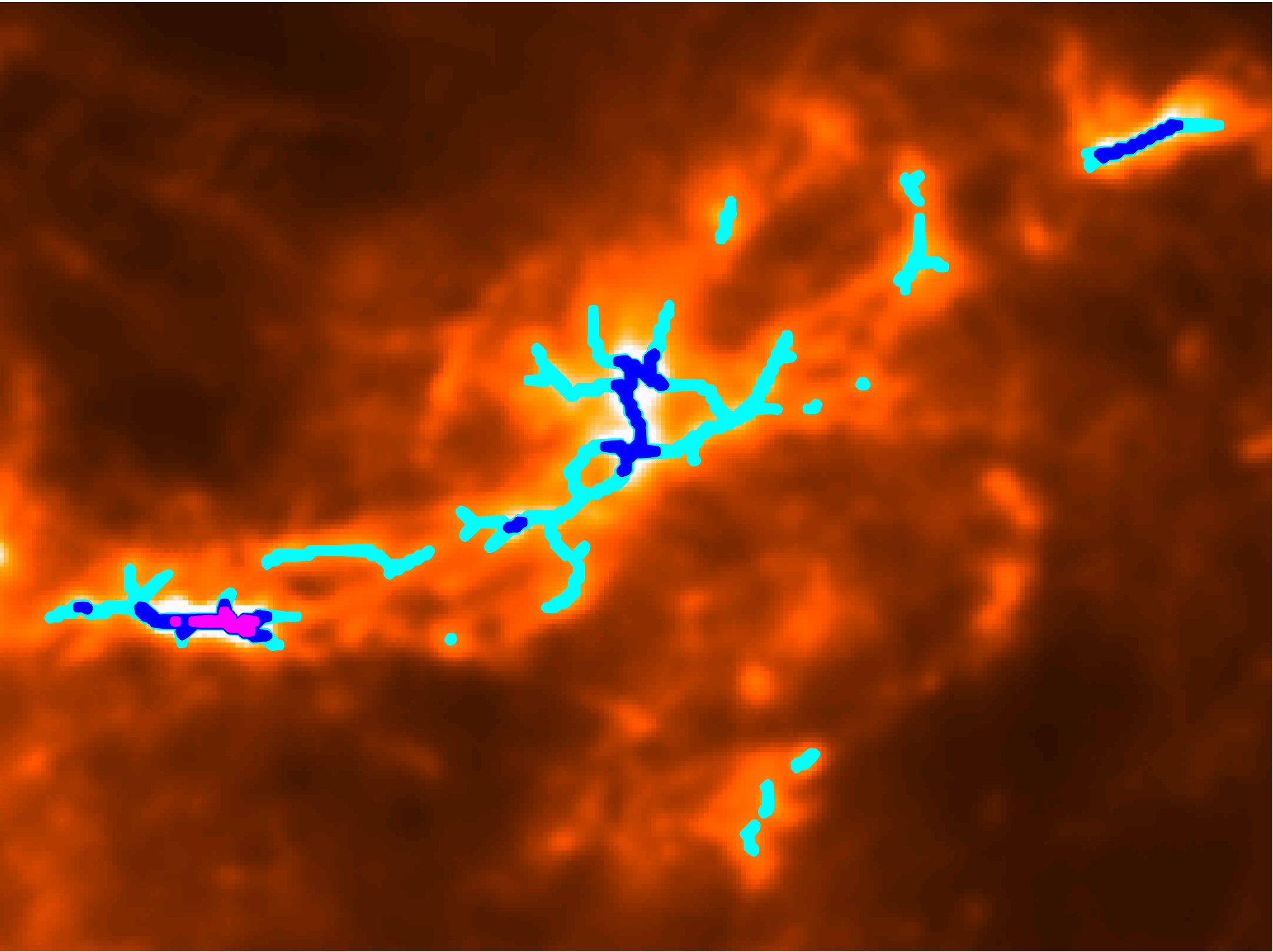}\\
\caption{The column density image of Vela~C including the filamentary structure detected by \disperse, with cyan, blue and magenta representing structures detected at \Av\ $>$\,25, 50 and 100\,mag, respectively. Ridges in this instance appear magenta in colour. Note that this figure is the same as presented in Figure \ref{fig:colden} with different colours only chosen here to highlight the crests. Bottom: zooms to South-Nest (left) and Centre-Ridge (right), the two most contrasting regions with respect to filamentary structure.
\label{fig:crestzooms}}
\end{center}

\end{figure*}

\subsection{Filaments and mass concentration}\label{sec:disperse}

To take a census of the filaments evident in the column density map and trace their crest\footnote{The crest is the maximum intensity profile along the filament.}, the discrete persistent structure extractor (\disperse)  was applied to the map. \disperse\ works \ch{on the principles of computational topology, namely the Morse theory, and 
 the} concept of persistence, to identify topological structure such as filaments, \ch{and connect their saddle-points to maxima with integral lines} 
\citep{sousbie11, sousbie11b}.

\ch{Figure \ref{fig:crestzooms} shows the crests identified from \disperse. } 
Here, `nest' depicts a disorganised network of filaments in contrast to a single dominating filament which we name `ridge', see Section \ref{sec:filam}. \ch{Since the Centre-Ridge appears to be more efficient in forming high-mass stars than the South-Nest (see Section \ref{sec:sf}) we utilise analytical tools in the following sections to help quantify this structural difference.}
Table \ref{tab:crest} gives the parameters of the filaments and their crest in each of the five sub-regions of Vela~C. \ch{This table shows that at low \Av\ (\Av~=~7\,--\,25\,mag), the filamentary networks are similar in terms of coverage (ratio of the number of crest points to the total number of points, column 3) in all sub-regions. The filamentary networks only differentiate at high \Av\ because only the Centre-Ridge and the South-Ridge have formed high-\Av\ structures. This can not be only a statistical result since the South-Nest, Centre-Nest and North sub-regions have slightly more pixels and more mass than the Centre-Ridge and South-Ridge (see Table \ref{tab:crest}).}
Above \Av\ $>$ 50 mag all filaments identified have supercritical masses
per unit length and are thus likely gravitationally unstable \citep[see discussion in][]{andre10}.
Fig. \ref{fig:colden} (right) shows that the Centre-Ridge has a filament whose outer radius is $\sim$1 -- 1.5\,pc, in marked contrast to those in the South-Nest which have radii $\sim$0.2 -- 0.3\,pc (see also Fig. \ref{fig:crestzooms}).

\begin{table*}
\caption{Table of parameters resulting from \disperse, relating to the filamentary structure seen in Vela~C. \ch{The numbers listed in the table are with respect to each region, which are sensitive to the region definition at \Av\ $>$\,7\,mag.}
Both of the regions that we have labelled as `ridges' are clearly distinguished from other regions with respect to their density. \label{tab:crest}}
\begin{tabular}{@{}lccccccccc@{}}
\hline
Region & Total & Tot nb.  & Crest   & Crest    & \multicolumn{2}{c}{Crest coverage at \Av\ $>$} & \multicolumn{3}{c}{Column density} \\
     & mass & pixels   &  pixels & coverage &  50\,mag & 100\,mag                         &
maximum & mean on crest & mean in whole region \\
       & $\times$10$^6$ \mstar\ & $\times$10$^4$ & $\times$10$^2$& \% & $\permil$  &$\permil$            &  $\times$10$^{22}$~cm$^{-2}$ & $\times$10$^{22}$~cm$^{-2}$ & $\times$10$^{22}$~cm$^{-2}$ \\
\hline
\hline
North      & 3.7 & 5.5 & 3.0 & 5.4 & 1.93 & 0    & 9.3  & 1.8 & 1.0 \\
Centre-Ridge & 3.6 &3.8 & 2.4 & 6.3 & 2.76 & 0.53 & 17.0 & 2.1 & 1.3 \\
Centre-Nest  & 3.3 &4.7 & 3.8 & 8.2 & 1.91 & 0    & 10.0 & 2.2 & 1.6 \\
South-Ridge  & 1.9 & 2.9 & 2.2 & 7.5 & 2.79 & 0.44 & 19.6 & 2.5 & 1.6 \\
South-Nest   & 4.2 &7.2 & 4.7 & 6.5 & 0.49 & 0    & 8.2  & 2.1 & 1.4 \\
\hline
\end{tabular}
\end{table*}

\begin{figure}
\begin{center}
\includegraphics[width=8.5cm]{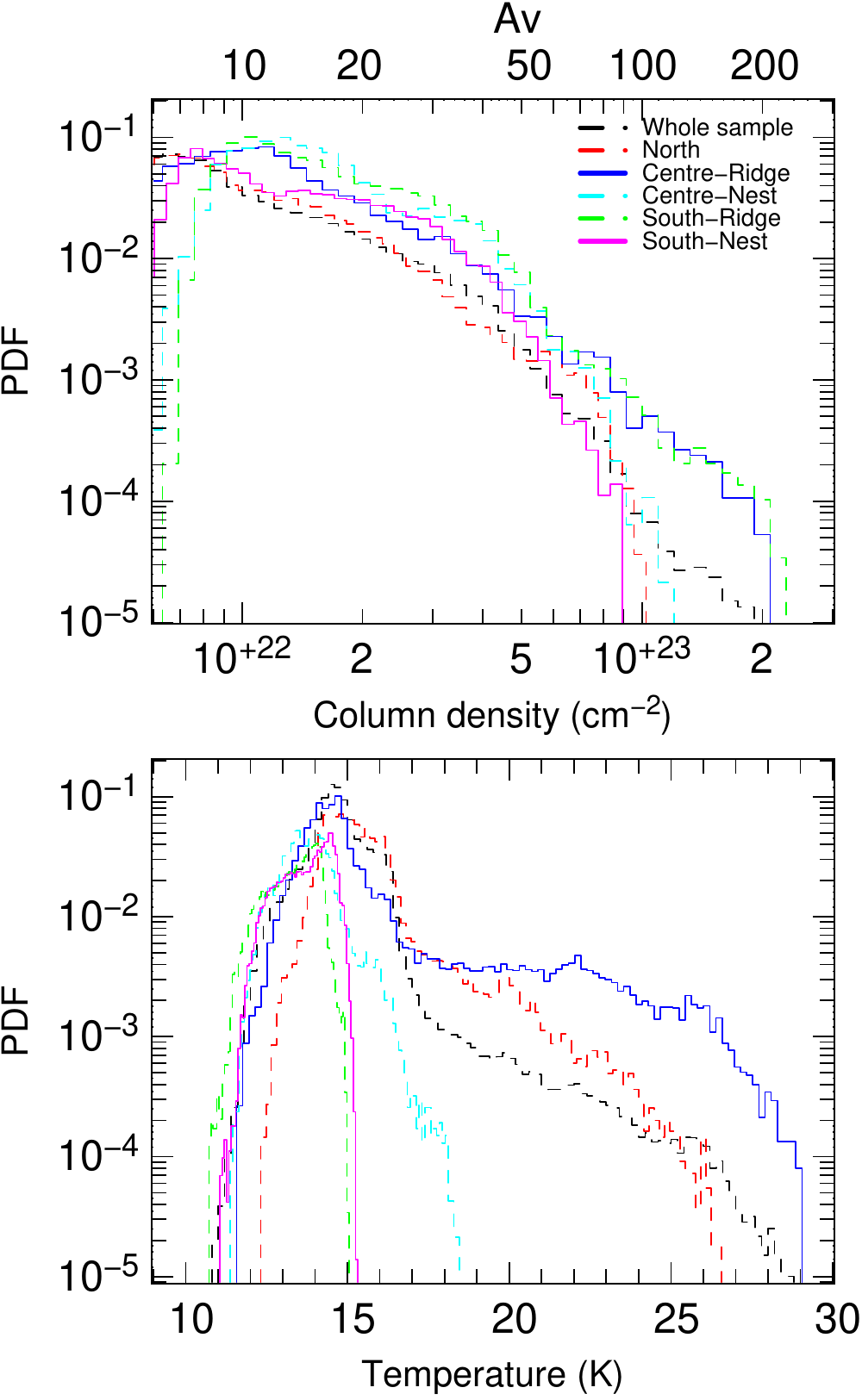}
\caption{Normalised column density (top) and temperature (bottom) probability distribution functions (PDFs) of Vela~C and its sub-regions. Note that the Centre-Ridge has a flatter column density slope than the South-Nest and a bimodal temperature distribution. \label{fig:pdf}}
\end{center}
\end{figure}

The column density and temperature probability distribution functions (PDFs) are presented in Fig. \ref{fig:pdf}.
Interestingly, the South-Ridge and South-Nest are the coolest sub-regions, with steep temperature PDF profiles spanning a small range ($<$16\,K). We focus here on the high-column density tails of the PDF, which are clearly flatter (see Table \ref{tab:crest}, Fig. \ref{fig:pdf}) in sub-regions with high-density filaments: Centre-Ridge and South-Ridge. \ch{Note that this behaviour is similar when considering only the pixels along the filaments (i.e. from \disperse) and the pixels forming these high-density tails directly correspond to structures which we call `ridges' (see Section \ref{sec:filam}).
}

\begin{figure}
\centering
\includegraphics[width = 8.5cm]{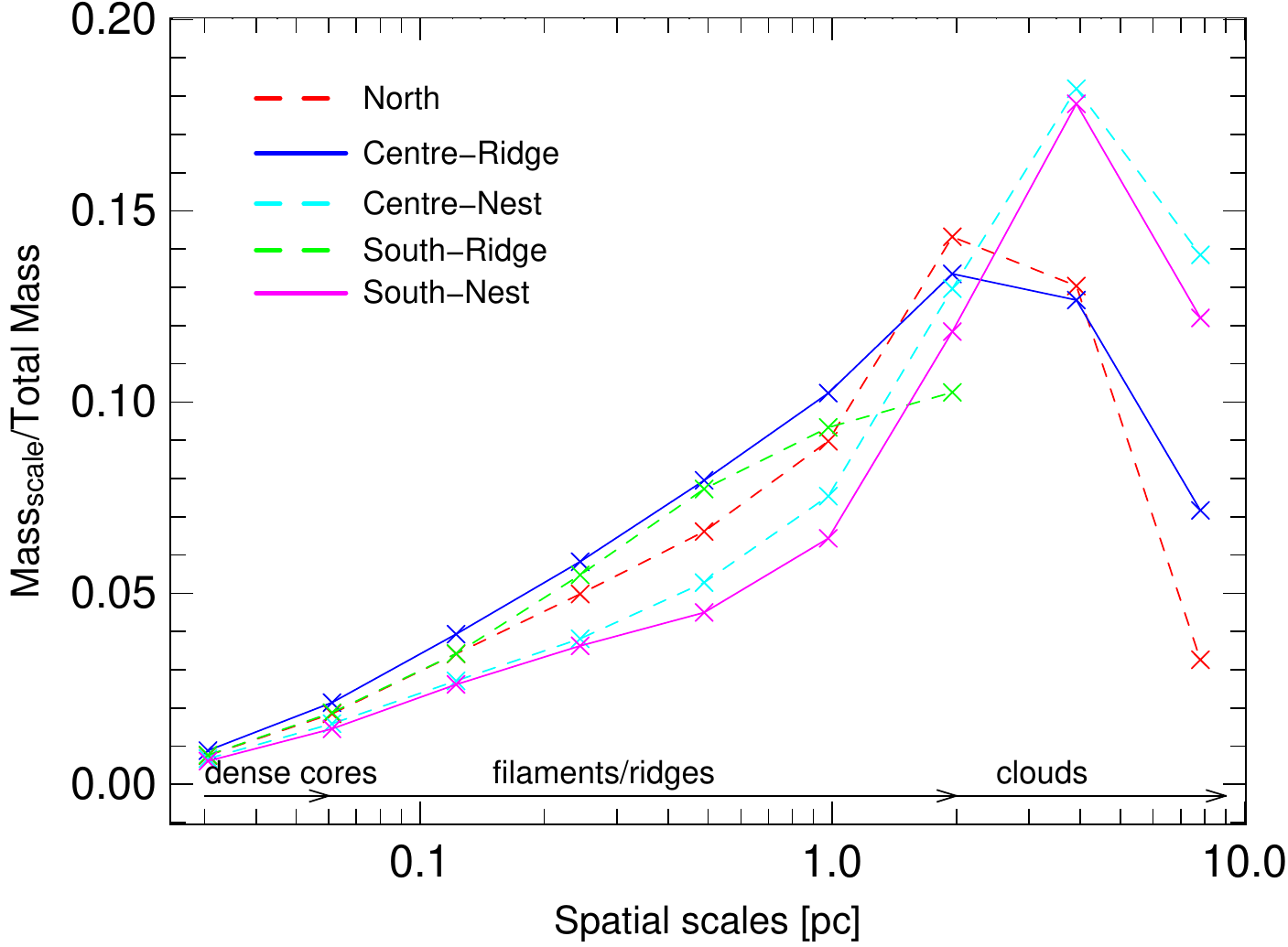}
\caption{MRA decomposition of the Vela~C column density map on nine scales.  Note the regular increase of mass ratio in the Centre-Ridge over all scales, compared with the South-Nest which has 
a shallow increase at $<$1\,pc and a sharper increase $>$\,1\,pc.
\label{fig:scales}} \end{figure}

\begin{figure}[]
\vspace{-1.5cm}
\centering
\includegraphics[width=6.3cm,angle=270]{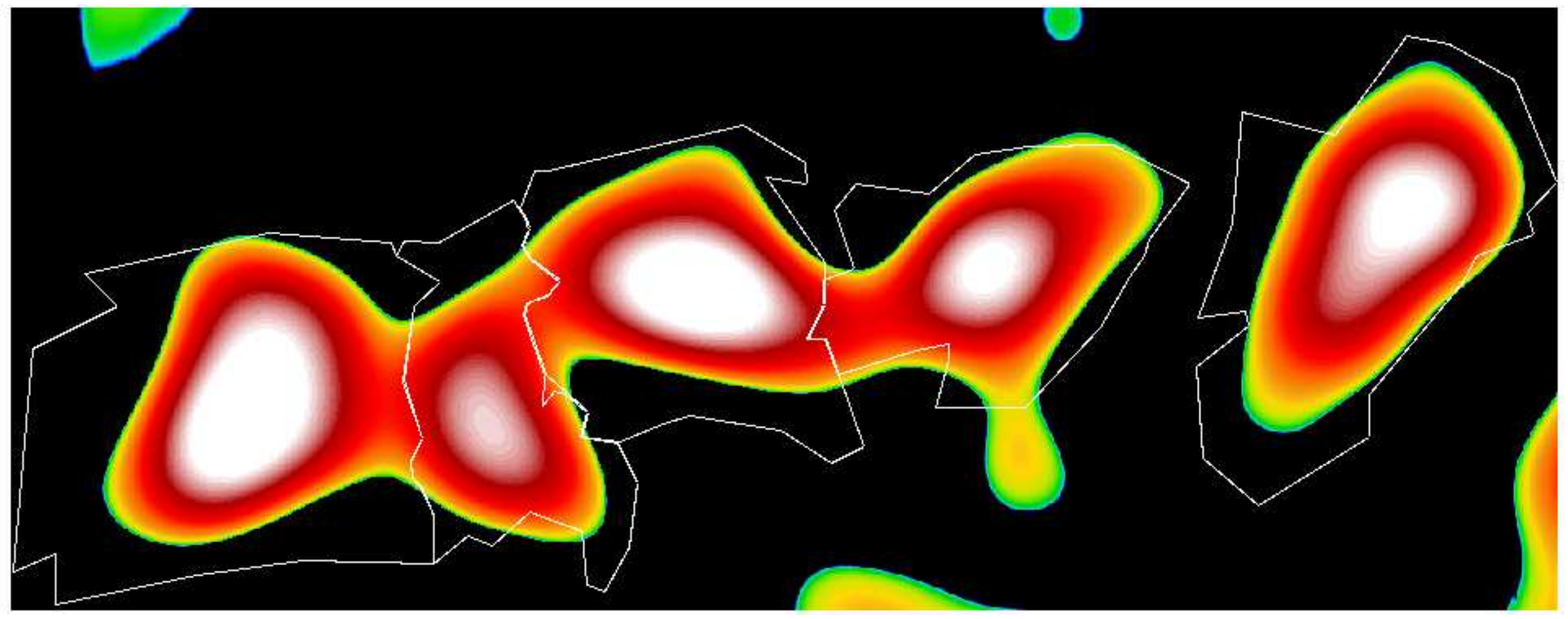}
\vspace{-1.5cm}
\caption{Scale 7 (2\,pc) of the MRA decomposition of Vela~C. The five sub-regions previously identified (Fig. \ref{fig:colden}) are confirmed by this analysis. \label{fig:mrfilter}}
\end{figure}

To follow the concentration of material within Vela~C, a multi-resolution analysis  (MRA) was performed on the column density map. MRA uses the wavelet transform to decompose a data series in a cascade from the smallest scales to the largest \citep[\emph{mrtransform},][]{starck94}. A MRA decomposition on nine scales, up to 38\arcmin, (i.e., $\sim$ 8\,pc) was made.
Figure \ref{fig:scales} shows the mass of cloud structures observed at certain scales after removing negative levels, divided by the total mass, as a function of their characteristic lengthscale. Small scales ($\lesssim$~0.07\,pc) represent dense cores, intermediate scales ($\gtrsim$~0.07\,pc and $\lesssim$~3\,pc) filaments and ridges while large scales ($\gtrsim$~3\,pc) represent the clouds.
The South-Nest and Centre-Nest regions have similar MRA profiles, as do the Centre-Ridge and South-Ridge regions from scales 1\,--\,6. \ch{While the mass concentration into dense cores ($\sim$ 2 -- 3\%) may be roughly equivalent for each sub-region (see Fig. \ref{fig:scales}), the concentration in filaments/ridges is higher and spans a greater scale range in the Centre-Ridge and South-Ridge than in the South-Nest: about 2/3 up to 1\,--\,2~pc 
 versus $\sim$ 
40\% up to $\sim$\,0.6~pc. Such a difference in mean size (average length and outer radius) is confirmed by a multi-resolution analysis of these filaments using a curvelet image built from the column density map (not shown here).}

\section{Low-mass vs. high-mass star formation in Vela~C}\label{sec:sf}

The Vela~C molecular complex offers a unique opportunity to study nearby high- and low-mass star formation sites at high spatial resolution with \herschel. Vela~C is segregated into five distinct sub-regions according to column density, each with different characteristics and filamentary structure. Relying on several methods, such as column density and temperature PDFs, \disperse, and MRA we characterise these sub-regions and relate them to their ability to form high-mass stars in the near future. Though the variation of their characteristics is probably continuous, for simplicity we hereafter discuss in detail the two most contrasting regions with respect to temperature and column density -  Centre-Ridge and South-Nest. 
We have identified high-mass ($>$\,8\,\mstar) compact sources with the highest signal to noise (S:N~$>$~50)  
in Vela~C, see Fig. \ref{fig:colden}. These \ch{13} objects have masses 
ranging from 20 -- 60\,\mstar\ and are thus capable of forming high-mass stars.
Based on their size of $\sim$0.03~--~0.2\,pc, these sources correspond to starless or protostellar dense cores.  \ch{Six} of them are found in the Centre-Ridge region, \ch{four} in the North, two in the Centre-Nest and one in the South-Ridge, while the South-Nest has none. \ch{Two of the sources in the North region are associated with the \hii\, region RCW\,34, and thus high-mass star formation. Of the remaining 11 high-mass sources identified, seven of them are found in high-density filaments, which we call `ridges' (see Section \ref{sec:filam}).
A census of the low- to intermediate-mass star formation in Vela~C is presented in a forthcoming paper by Giannini et al., in prep.}

The Centre-Ridge sub-region has already formed a cluster of massive stars, which is heating the low- to medium-density (\Av\ $> $10~mag) molecular cloud next to it. The Centre-Ridge is also home to a radio continuum source
\citep[][08576-4334]{walsh98}, discussed in detail by Minier et al. (in prep).
The temperature PDF of the Centre-Ridge (and North) displays a bimodal distribution, affirming the presence of a hot pocket of gas (RCW\,36) around the OB cluster, and a cold filament which has not yet been impacted, at least in its centre,
 by this cluster. The warmer component arises solely from the \hii\ regions present, as expected (Fig.~\ref{fig:dusttemp}). In the absence of this warmer component, the Centre-Ridge displays a temperature PDF profile similar to that of the other regions, though extending to slightly warmer temperatures.
We measure a temperature of $\sim$\,15 -- 18\,K for the \ch{inner part of the} filament, which is slightly higher than the $\sim$10 -- 12\,K found in other parts of Vela~C.
\ch{Perpendicular to this filament we measure a density profile close to that of the classical  r$^{-2}$ profile. Stellar winds or UV radiation from the nearby OB cluster
 are expected to steepen a density profile \citep[e.g.][]{lefloch94, hennebelle03}.}
The density profile (Fig. \ref{fig:colden}, right) of this filament does not show such strong steepening \ch{suggesting that it is not yet affected on large scales by} the proximity of the OB cluster.
 We have thus found massive stars forming in a filamentary structure from 
initial temperature conditions not very different to that found in low-mass star-forming sub-regions like the South-Nest.

\ch{Each of the sub-regions of Vela C contains roughly the same mass (1.9 -- 4.2 $\times$10$^6$ \mstar) and, assuming a constant star-formation efficiency, should form approximately the same number of massive stars.} In contrast, the Centre-Ridge contains more massive and thus denser cores than the South-Nest that may correspond to the next generation of high-mass stars.

\section{Filaments vs. ridges}\label{sec:filam}
 
In the column density image (Fig. \ref{fig:colden} or \ref{fig:crestzooms}) there is clear filamentary structure throughout the entire Vela~C complex, with the brightest and densest filaments forming a ridge-like structure running through the centre of the map (i.e. the Centre-Ridge).  Above an \Av\ $>$ 50 to 100 mag, there is little dense structure detected by \disperse\ (see Fig. \ref{fig:crestzooms}),  whilst above 100\,mag only two filamentary structures are detected.
We refer to these two notable filamentary structures, which reach a column density greater than  10$^{23}$\,cm$^{-2}$, as `ridges'. \ch{\citet{krumholz08} suggest the existence of a column density threshold for massive stars, where only clouds which have a column density $>$\,1\,g/cm$^{2}$ \ch{(corresponding to N$_{H_2}$ $\sim$ 3\,$\times$\,10$^{23}$ cm$^{-2}$)} can avoid fragmentation and form massive stars.} Whilst our results are consistent with the
high-mass star formation predictions of \citet{krumholz08}, their models are based on a quasi-static view of cloud formation and
our data do not suggest the presence of a definitive threshold between high- and low-mass star-forming sites.
 In Vela~C the main ridge is a massive supercritical filament that differs from a filament forming low-mass stars in terms of its larger width (with a statistical distance between filaments $\sim$1\,pc compared with 0.2\,pc) and more complex structure (Minier et al., in prep) leading to a wider FWHM when fit with a single Gaussian ($\sim$0.3\,pc).
 Two sub-regions (Centre-Ridge and South-Ridge) contain a ridge,  (see Fig. \ref{fig:colden}), which is 
the dominant density structure of each of these regions. In contrast, the South-Nest region of Vela~C houses a network of weaker filaments. 
 These filaments are less ordered than in the other regions of the complex, suggesting governance or formation through turbulence - see Fig. \ref{fig:crestzooms}.

The few \Av\ $>$ 100 mag $\sim$ 3\,pc filamentary structures identified in Fig.~\ref{fig:hipe} (500\,\microm) have no equivalent in low-mass star-forming regions \citep[e.g.][]{arzoum11}. They correspond to the dominating filaments that we qualified as 
ridges and could be the potential sites of high-mass star formation. Such ridges have been identified in other star-forming regions such as DR\,21 in Cygnus\,X \citep{schneider10b} \ch{or W43-main in W43 \citep{quang11}. Such self-gravitating, massive cloud structures with large aspect ratios are difficult to form in numerical simulations of molecular clouds in static boxes but a few good candidates may be seen in colliding flows simulations \citep[e.g.][]{heitsch09}}.

\section{Stellar content relative to cloud structure}\label{sec:relative}

Star formation is controlled by the interplay between gravity and turbulence. On large scales, turbulence provides stability whilst on smaller scales it can create local high density regions where self-gravity can
take hold.
When a molecular cloud is mostly shaped by turbulence, the mass of three-dimensional structures should be directly proportional to the radius squared (M $\propto$ r$^2$) \citep{larson81} and when gravity is the dominant factor shaping the molecular cloud, the relation is M~$\propto$~r \citep{bonnor56}. At small and intermediate scales 
($<$ 0.06 - 1\,pc) the MRA profiles of Centre-Ridge and South-Nest (Fig. \ref{fig:scales}) are consistent with a relation for gravitationally-bound (Bonnor-Ebert-like) filaments, while at scales $>$\,1\,pc, 
the behaviour of the South-Nest (and Centre-Nest) approaches that of a turbulent medium.
A MRA approach (Fig.~\ref{fig:scales}) reveals a stronger
concentration of material 
in the filaments/ridges of the Centre-Ridge than in the South-Nest. The mass distribution of the Centre-Ridge, i.e., well ordered dense structure such as
 the ridge and several high-mass protostars, suggests that it is governed by gravity rather than turbulence.

\ch{The column density probability distribution function has been widely used by observers and theoreticians to statistically characterise the cloud structure as a function of the star-formation activity. \citet{kain09} find that the shape of their observed column density distributions changes with the star formation status, clouds with active star formation develop a power-law tail at high column densities, in agreement with some numerical simulations including gravity \citep[e.g.][]{bp11}. 
Our PDF analysis focuses on high-column density regions, above the \Av\ $\sim 8-10$~mag mentioned for the beginning of such a power-law tail. The striking difference between the flat tail of the Center-Ridge and the steepening of the South-Nest reflects the presence or absence of a ridge. According to some models, tails are indicative of gravity-dominated regions \citep{klessen00} but others suggest that they are the signature of stronger cloud evolution \citep{cho11}.}

Above \Av\ $>$ 30\,mag, the Centre-Ridge reaches a higher column density with a flatter PDF than the South-Nest.
 A flatter slope is expected for coherent structures created via constructive large-scale flows rather than small scale turbulence \citep[e.g.][]{federrath10b}. A higher column density is achieved when the compression of material is stronger and thus the velocity difference of converging flows is probably higher. \cha{The PDFs of the filament crest pixels (from \disperse, not shown) mirror the PDFs of the corresponding sub-regions (Fig.~\ref{fig:pdf}) derived from the column density map.}
The steeper PDF and the smaller concentration of material in the South-Nest, with respect to the Centre-Ridge, suggests this region to be more turbulent.

\vspace{0.3cm}

\section{Summary}

\ch{We present 3 square degree \herschel\ maps of Vela C in the far-infrared and submillimetre, using the PACS and SPIRE instruments. The Vela C molecular complex houses both hot evolved objects such as RCW\,36, and more cold diffuse filamentary structure. At \cha{\Av\ = 7 mag}, the Vela C complex segregates into five distinct sub-regions, which is confirmed by a multiresolution analysis. Using probability distribution functions and a multiresolution analysis}
both of the South-Nest and Centre-Ridge sub-regions, \ch{the two most contrasting sub-regions}, appear to be forming stars though the Centre-Ridge contains more massive protostellar candidates than the South-Nest.
 The above structure analysis suggests that in Vela~C high-mass star formation proceeds preferentially in the very high column density dominant filaments we called ridges
(reaching N$_{H_2}$~$\sim$~10$^{23}$~cm$^{-2}$) which may result from the constructive convergence of flows. Such processes have been advocated for the formation of high-mass stars \citep[cf][]{schneider10b, csengeri11}.

\begin{acknowledgements}
T.H. is supported by a CEA/Marie-Curie Eurotalents Fellowship. Part of this work was supported by the ANR (\emph{Agence Nationale pour la Recherche}) project `PROBeS', number ANR-08-BLAN-0241. This work has made use of the TOPCAT \citep{taylor05} and Yorick freeware packages (see  http://www.starlink.ac.uk/topcat/, http://yorick.sourceforge.net/). Fig. \ref{fig:hipe} was generated using APLpy, an open-source plotting package for Python hosted at http://aplpy.github.com. \ch{We would like to thank the anonymous referee, whose comments contributed to improve the manuscript.}

SPIRE has been developed by a consortium of institutes led by
Cardiff Univ. (UK) and including Univ. Lethbridge (Canada);
NAOC (China); CEA, LAM (France); IFSI, Univ. Padua (Italy);
IAC (Spain); Stockholm Observatory (Sweden); Imperial College
London, RAL, UCL-MSSL, UKATC, Univ. Sussex (UK); Caltech, JPL,
NHSC, Univ. Colorado (USA). This development has been supported
by national funding agencies: CSA (Canada); NAOC (China); CEA,
CNES, CNRS (France); ASI (Italy); MCINN (Spain); SNSB (Sweden);
STFC (UK); and NASA (USA).

PACS has been developed by a consortium of institutes led by MPE (Germany) and including UVIE (Austria); KU Leuven, CSL, IMEC (Belgium); CEA, LAM (France); MPIA (Germany); INAF-IFSI/OAA/OAP/OAT, LENS, SISSA (Italy); IAC (Spain). This development has been supported by the funding agencies BMVIT (Austria), ESA-PRODEX (Belgium), CEA/CNES (France), DLR (Germany), ASI/INAF (Italy), and CICYT/MCYT (Spain).

\end{acknowledgements}

\bibliographystyle{aa} 

\Online

\begin{appendix} \centering
\section{Online Material}

\begin{figure*}[!ht]
\centering
\includegraphics[height=0.4\hsize]{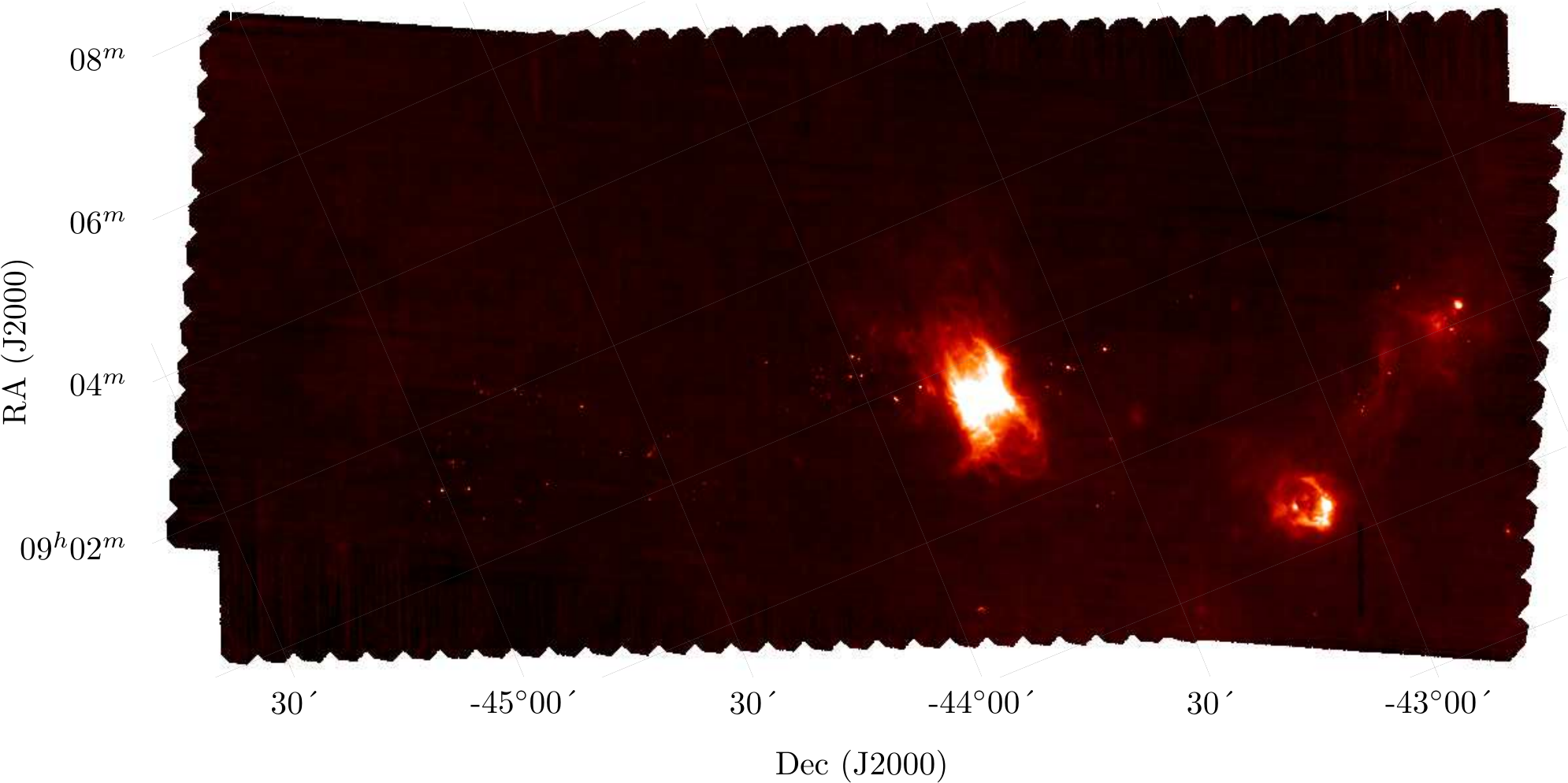}
\end{figure*}

\begin{figure*}[!ht]
\centering
\includegraphics[height=0.4\hsize]{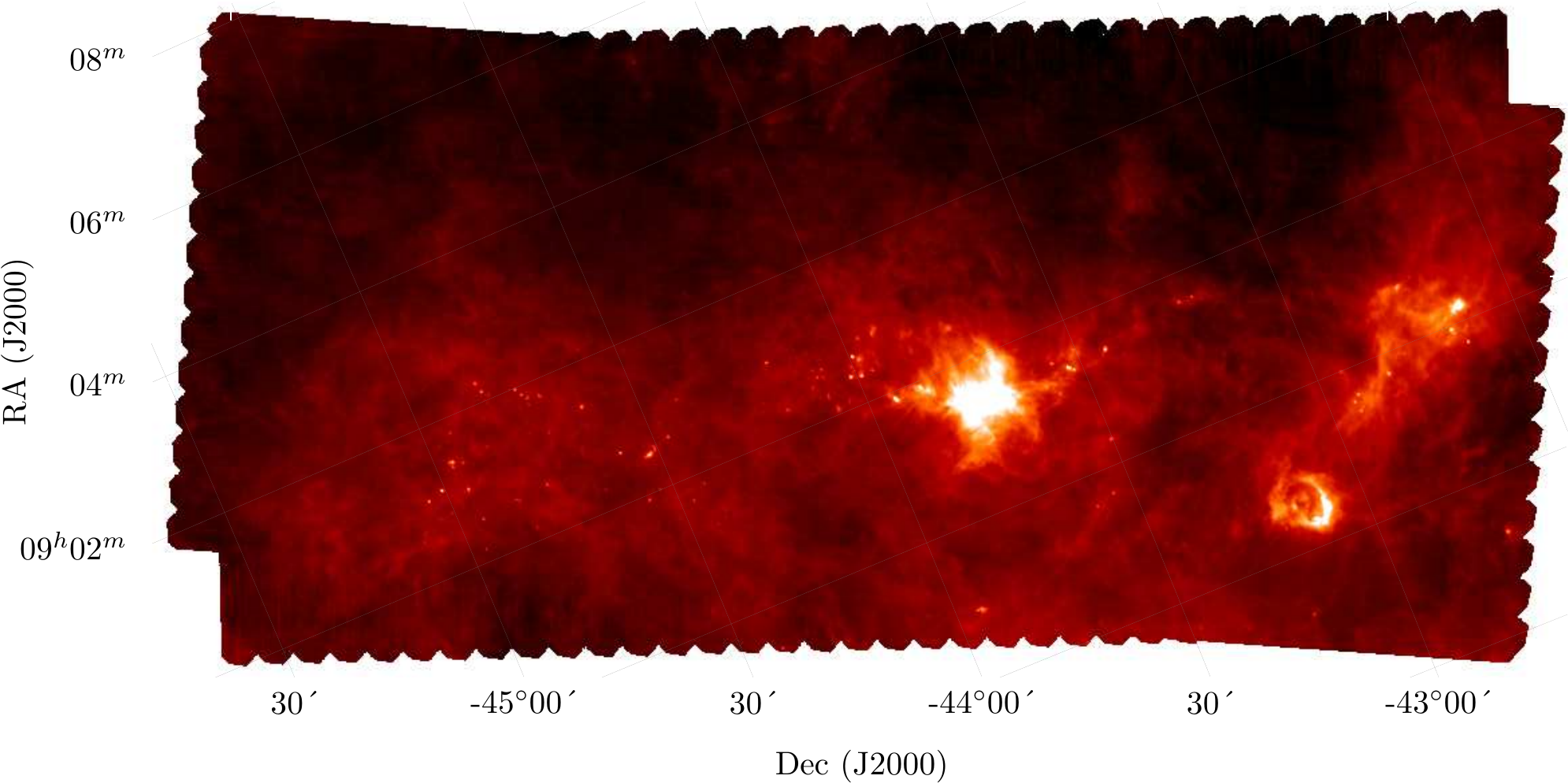}
\end{figure*}

\begin{figure*}[!ht]
\centering
\includegraphics[height=0.4\hsize]{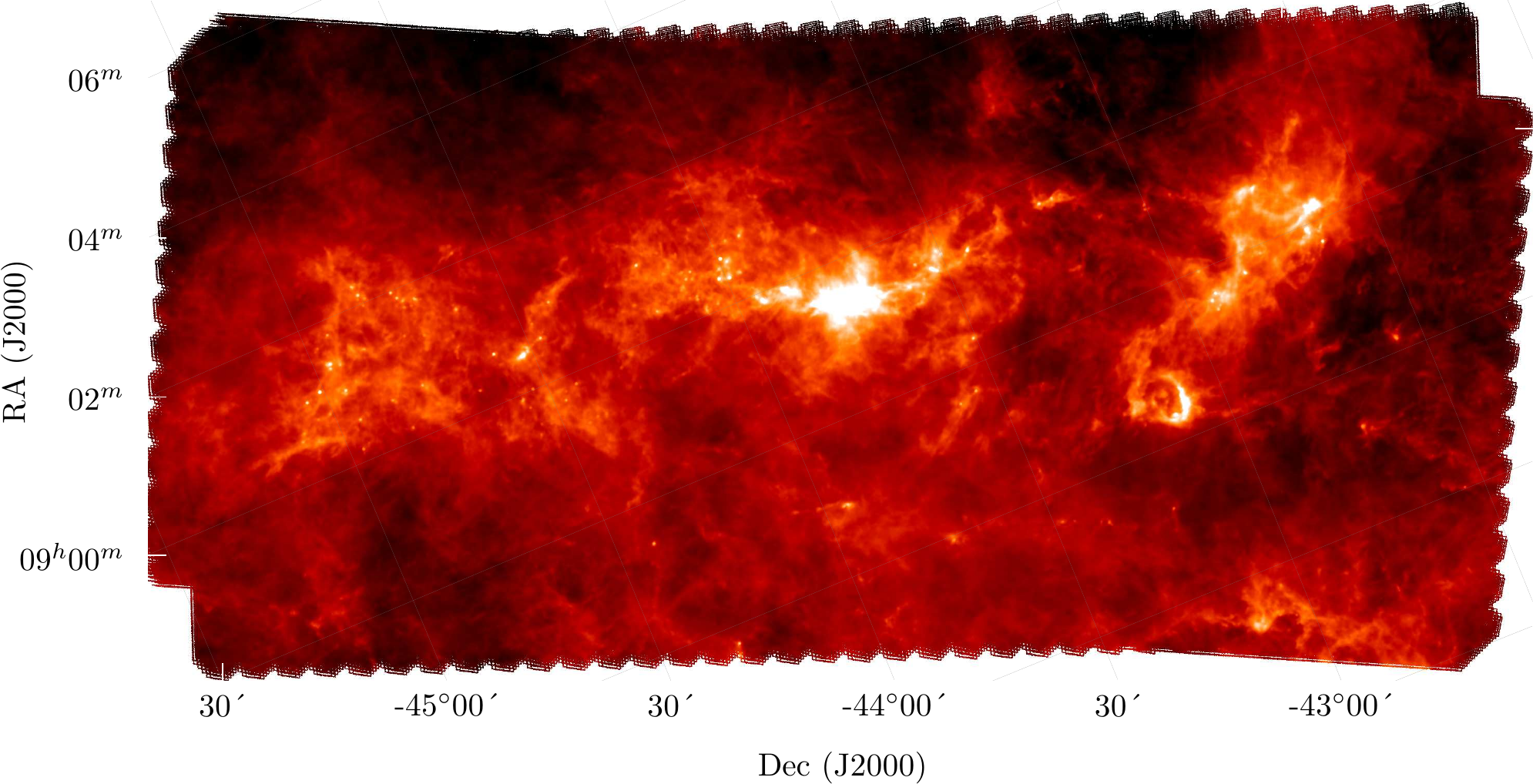}
\end{figure*}

\begin{figure*}
\centering
\includegraphics[height=0.4\hsize]{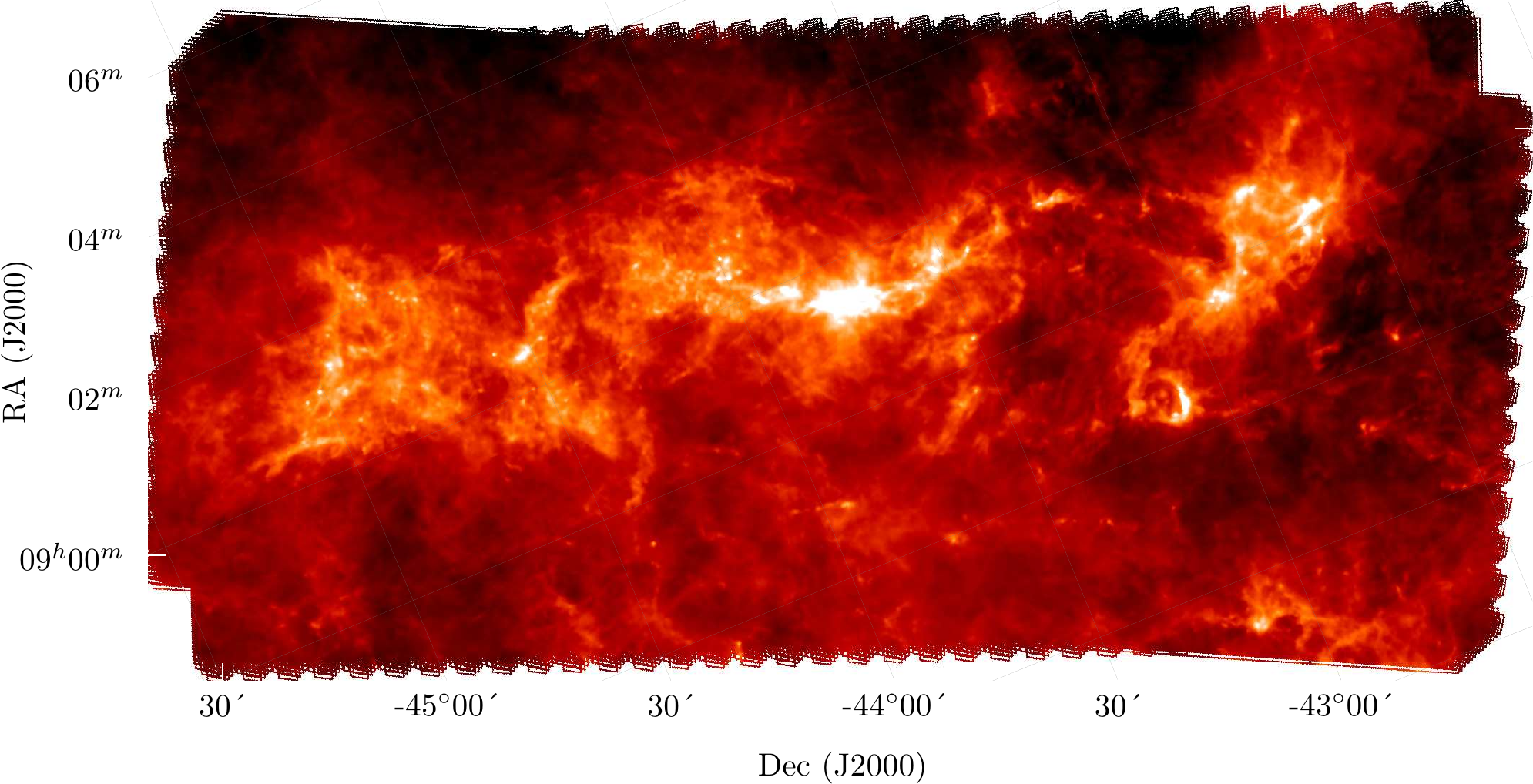}
\end{figure*}

\begin{figure*}
\centering
\includegraphics[height=0.4\hsize]{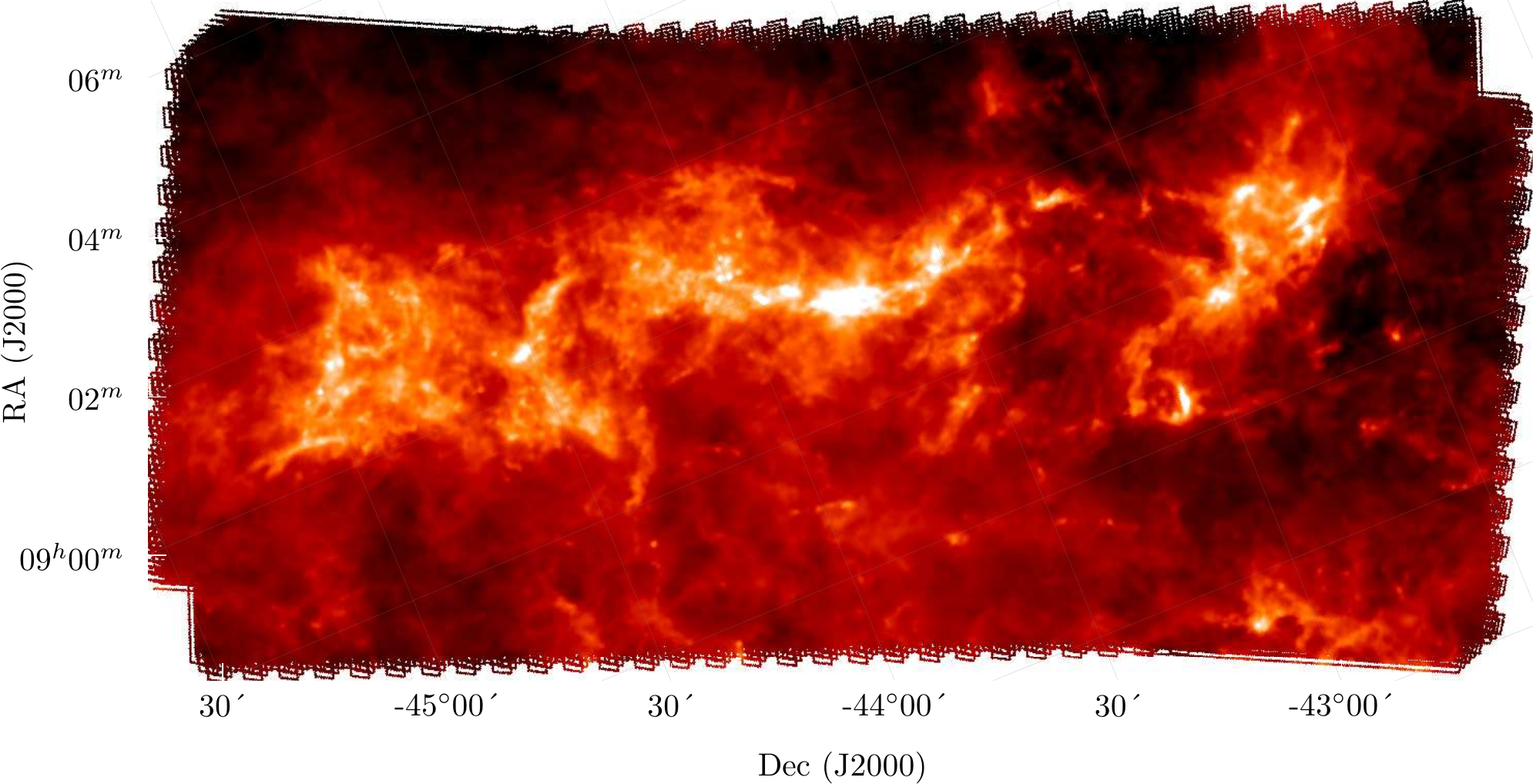}
\caption{The 5 \herschel\, wavebands of Vela~C, from top to bottom, 70, 160, 250, 350 and 500\,\microm. \ch{The observed rms level of cirrus noise in Vela~C is $\sim$ 6\,mJy and 20\,mJy,  at $\lambda$\,=70 and 160\,\microm, respectively and $\sim$\,200\,--\,500\,mJy at the SPIRE 250, 350 and 500\,\microm\ bands.}
 At shorter wavelengths only those warmer objects, such as protostars and \hii\ regions are seen. At longer wavelengths \herschel\ detects cold, deeply embedded filaments and the progenitors of high-mass stars. 
\label{fig:hipe}}
\end{figure*}

\end{appendix}
\end{document}